\documentclass{PoS}

\usepackage[font=small,labelfont=bf]{caption} 

\title{Single top-quark production with the Matrix Element Method in
  next-to-leading order accuracy}

\ShortTitle{Single top with MEM at NLO}

\author{\speaker{T. Martini} and P. Uwer\\
  Humboldt-Universit{\"a}t zu Berlin, Institut f{\"u}r Physik,
  Newtonstra{\ss}e 15, 12489 Berlin, Germany\\
  E-mail: \email{Till.Martini@physik.hu-berlin.de},
  \email{Peter.Uwer@physik.hu-berlin.de}}

\abstract{Single top-quark production offers a
  unique laboratory for precision tests of the Standard Model and
  searches of possible extensions. Furthermore, assuming the Standard
  Model, single top-quark production can be used to determine
  top-quark related couplings. For precise determinations of
  parameters like the electroweak gauge couplings or the mass of the
  top quark, efficient, unbiased, and theoretically unambiguous analysis
  methods are needed.  Within this context, the Matrix Element Method
  (MEM) has been established in hadron collider analyses due to its
  possibility to top out at utilising the information available in
  experimental data.  However, so far it has mostly been applied in
  Born approximation only.  We discuss the extension to
  next-to-leading order (NLO) accuracy. As a necessary prerequisite we
  introduce an efficient method to calculate NLO QCD weights for jet
  events. As proof of concept and representative example we use the
  MEM at NLO to reproduce the top-quark mass in a toy experiment where
  we treat single top-quark events generated at NLO accuracy as
  pseudo-data. This is the first application of the MEM at NLO
  accuracy to the hadronic production of jets originating from coloured
  final state partons. We observe that analysing NLO events with Born
  likelihoods can introduce a pronounced bias in the extracted mass
  which would require significant calibration with associated
  uncertainties. Although we focus on parameter determinations, the
  methods presented here can also be used to search for new physics
  using likelihood ratios.}

\FullConference{XXV International Workshop on Deep-Inelastic
  Scattering
  and Related Subjects\\
  3-7 April 2017\\
  University of Birmingham, UK}

\begin{document}

\section{The MEM in a nutshell}
For the collision $A+B\rightarrow a_1(p_1)+a_2(p_2)+...+a_n(p_n)$ the
differential cross section ${d\sigma\over dR_n(p_1,\ldots,p_n)}\propto
|\mathcal{M}(p_1,\dots,p_n)|^2dR_n$ is a measure for the probability
to observe the final state in the infinitesimal phase space region
$dR_n=(2\pi)^4\delta(P-\sum\limits_{i=1}^{n}
p_i)\prod\limits_{j=1}^{n}{d^3p_j\over(2\pi)^32E_j}$ located at
$(p_1,\dots,p_n)$. By collecting generic partonic final state
variables (e.g. $E_i, \eta_j,\dots$) in $\vec{y}$ we can generalise
the differential cross section to ${d\sigma\over dR_n(p_1,\ldots,p_n)}\rightarrow{d\sigma_n\over d\vec{y}}$. Experimentally, only
hadronic variables $\vec{x}$ instead of partonic variables $\vec{y}$
are observed. The correspondence between $\vec{y}$ and $\vec{x}$ is
given by the so-called Transfer Function $W(\vec{x},\vec{y})$ which
models the probability to observe a partonic event $\vec{y}$ as a
hadronic event $\vec{x}$.  These Transfer Functions have to be
determined by experiments through detector simulations. In the
following, we will set them to $\delta$-functions as a first
approximation assuming an ideal detector. Non-trivial transfer
functions require additional integrations over the partonic variables
but do not pose a principle restriction for the method presented
here. With these deliberations we can calculate a model dependent
likelihood for having measured a specific event sample $\{\vec{x}_i\}$
as the product of the individual probabilities for each event in 
the sample
\begin{displaymath}
  \mathcal{L}({\omega})=\prod\limits_{i}\frac{1}{\sigma}({\omega})
  \int d\vec{y}{d\sigma({\omega})\over d\vec{y}}
  \underbrace{W(\vec{x_i},\vec{y})}_{=\delta(\vec{x_i}-\vec{y})}
  =\prod\limits_{i}{1\over\sigma({\omega})}{d\sigma({\omega})\over d\vec{x_i}}.
\end{displaymath}
This likelihood is a function of model parameter(s) ${\omega}$
entering the cross section calculation within a given model.
Maximising the likelihood for the event sample at hand with respect to
${\omega}$ yields an estimator ${{\widehat{\omega}}}$ for the model
parameter(s):
$\mathcal{L}(\widehat{{\omega}})=\sup_{{\omega}}\mathcal{L}({\omega})$.
Because all information from the event is used in the matrix element
when evaluating the likelihood the MEM ideally results in the most
efficient estimator. Pioneered at the Tevatron (see e.g. ref.
\cite{Abbott:1998dn,Abazov:2004cs,Abulencia:2006mi}) the MEM has been
widely used since. However, the experimental analysis is restricted to
Born approximation so far.
\section{Extending the MEM to NLO}
In the past some effort was put into including higher order
corrections into the MEM: The effect of QCD radiation was studied in
ref. \cite{Alwall:2010cq}. The hard Born matrix element is combined
with a parton shower in ref. \cite{Soper:2014rya}. Full NLO
corrections to the production of uncoloured objects were introduced in
ref. \cite{Campbell:2012cz} and applied in ref.
\cite{Campbell:2013hz}. Ref. \cite{Campbell:2013uha} investigates the
inclusion of hadronic jet production by mapping NLO and LO jets with a
boost along the beam axis to balance the transverse momentum. A
complete algorithm based on modified $(3\rightarrow 2)$ clustering
prescriptions to extend the MEM to NLO for arbitrary initial and final
states was published in ref. \cite{Martini:2015fsa} . Ref.
\cite{Baumeister:2016maz} suggests to retain the $(2\rightarrow 1)$
clustering prescriptions by numerically solving a system of non-linear
equations.

When calculating cross sections new features enter at NLO: Infrared
(IR) divergences in virtual and real contributions have to be mutually
cancelled and the $n+1$ particle phase space of the real corrections
introduces a non-trivial mapping of parton momenta $(p_1,\dots
,p_{n+1})$ to the jet momenta $(\widetilde{J}_1(p_1,\dots
,p_{n+1}),\dots,\widetilde{J}_n(p_1,\dots ,p_{n+1}))$. 
Extending the likelihood to NLO accuracy using 
\begin{displaymath}
{ \mathcal{L}^{\mbox{\scriptsize NLO}}({\omega})=\prod\limits_{i}\frac{1}{\sigma^{\mbox{\scriptsize NLO}}_{\mbox{\scriptsize$n$-jet}}({\omega})}\left({\frac{d\sigma^{\mbox{\scriptsize B+V}}_{n\rightarrow\mbox{\scriptsize$n$-jet}}({\omega})}{dR_n(J_1,\ldots,J_n)}}+{\frac{d\sigma^{\mbox{\scriptsize R}}_{n+1\rightarrow\mbox{\scriptsize$n$-jet}}({\omega})}{dR_n(\widetilde{J}_1,\ldots,\widetilde{J}_n)}}\right)\Big|_{\small{\mbox{event } i}}}  
\end{displaymath}
where $d\sigma^{\mbox{\scriptsize B+V}}_{n\rightarrow\mbox{\scriptsize $n$-jet}}$
($d\sigma^{\mbox{\scriptsize R}}_{n+1\rightarrow\mbox{\scriptsize $n$-jet}}$)
denotes the sum of Born and virtual (real) contributions,
three requirements have to be fulfilled:
\begin{enumerate}
\vspace{-1.4ex}
\item Both contributions are separately IR divergent. To ensure a
  point-wise cancellation within phase space both contributions have
  to be evaluated for the same jet momenta: $J_i=\widetilde{J}_i$.
  \vspace{-1.4ex}
\item In the real contribution the clustering of $n+1$ partons to $n$
  jets introduces $\delta$-functions
  $\delta(J_i-\widetilde{J}_i(p_1,\dots ,p_{n+1}))$ in the phase space
  integration which render a straight-forward numerical integration
  problematic. 
  \vspace{-1.4ex}
\item To evaluate the Born and virtual matrix elements for the jet
  momenta the clustered jets have to be on-shell: $J^2_i=m^2_i$ and
  respect momentum conservation: $J_1+\dots+ J_n=p_1+\dots+p_{n+1}$ at
  the same time.  \vspace{-1.4ex}
\end{enumerate}
To meet all three requirements at once we propose to replace the
common $(2\rightarrow 1)$ clustering prescriptions in jet algorithms
with $(3\rightarrow 2)$ clustering prescriptions inspired by the
Catani-Seymour dipole subtraction method (see ref.
\cite{Catani:2002hc}). Using these modified jet algorithms allows to
factorise the phase space $dR_{n+1}(p_1,...,p_{n+1})=
dR_n(\widetilde{J}_1,...,\widetilde{J}_n)dR_{\textrm{\scriptsize
    unres}}(\Phi)$ and offers the possibility to consistently define a
differential jet cross section (jet event
weight) at NLO accuracy :
\begin{equation}\label{eq:master}
  \frac{d\sigma^{\mbox{\scriptsize NLO}}_{\mbox{\scriptsize
        $n$-jet}}(\omega)}{dR_n(J_1,\ldots,J_n)}
  ={\frac{d\sigma^{\mbox{\scriptsize B+V}}_{n\rightarrow\mbox{\scriptsize $n$-jet}}(\omega)}{dR_n(J_1,\ldots,J_n)}}
  +{{\int\!\!dR_{\mbox{\scriptsize
          unres}}(\Phi)}}\left.{\frac{d\sigma^{\mbox{\scriptsize R}}_{n+1\rightarrow\mbox{\scriptsize $n$-jet}}(\omega)}{dR_{n+1}(p_1,\ldots,p_{n+1})}}
\right|_{p_i=p_i(J_1,\ldots,J_n,\Phi)}.
\end{equation}
For details we refer to \cite{Martini:2015fsa}. To cancel IR
divergences in both terms, we employ the Phase Space Slicing (PSS)
method. The NLO jet event weight can be used to
generate unweighted NLO jet events or to construct the NLO likelihood
for the MEM.
\section{Validation}
To validate the real phase space generation and the generation of
unweighted events we reproduce NLO jet distributions calculated using
a conventional parton level Monte-Carlo generator with a
$(3\rightarrow 2)$ jet algorithm. The $s$- and $t$-channel production
of single top quarks in association with a light jet at the LHC
($\sqrt{s}=13$ GeV) is studied: $pp\rightarrow t j (X)$. We study the
exclusive (additional jet activity $X$ is vetoed) and the inclusive
(additional jet activity
$X$ is allowed) case. For simplicity we do not consider the decay of
the top quark but treat them as tagged top jets. Details on the
calculation using Phase Space Slicing to cancel the IR divergences 
can be found in ref. \cite{Cao:2004ky}.\\
\begin{minipage}{\linewidth}
\vspace{2ex}
  \centering
    \includegraphics[width=0.49\textwidth]{{{%
          sgtsKT3-2ycut30compmnspecETA1-50x1e7_pos-crop}}}
    \includegraphics[width=0.49\textwidth]{{{%
          sgttKT3-2ycut30compmnspecincE2-114evts_pos-crop}}}

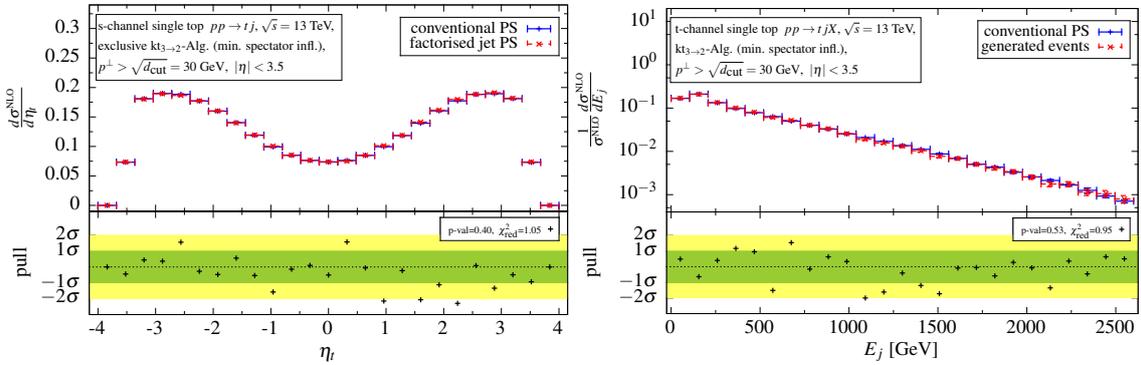
\captionof{figure}{Validation of phase space generation (left) and
  event 
  generation (right).}
\label{fig:valid}
\vspace{2ex}
\end{minipage}
The left hand side of fig. \ref{fig:valid} shows the comparison of the
pseudo rapidity distribution of the top-tagged jet in the exclusive
$s$-channel obtained using a conventional parton-level MC (solid blue)
and the method outlined in eq.~\ref{eq:master} (red dashed). The right
hand side compares a histogram of the energy of the light jet filled
with unweighted inclusive t-channel NLO events (red dashed) generated
using again eq.~\ref{eq:master}. These results are compared to the
distribution obtained from a conventional parton-level MC (solid blue).
In both cases, we find perfect agreement (see pull distribution at the
bottom of the plots). Analogous comparisons of other observables for
the exclusive and inclusive $s$- and $t$-channel show similar results and
can be found in ref. \cite{Martini:2017}.
\section{Application}
The NLO jet event weight eq.~\ref{eq:master} can be used in the MEM at 
NLO to recover the input value of the top mass 
$m^{\scriptsize \mbox{true}}_t=173.2$ GeV from the generated NLO jet 
events $\vec{x}_i=(\eta_{t},E_{j}, \eta_{j}, \phi_{j})_i$ e.g.:
\begin{displaymath}
  \mathcal{L}^{\scriptsize
     \mbox{NLO}}(m_t)=\prod\limits_{i}^N\mathcal{L}^{\scriptsize\mbox{NLO}}(\vec{x}_i|m_t)=\left(\frac{
       1}{\sigma^{\scriptsize \mbox{NLO}}(m_t)}\right)^N
   \prod\limits_{i=1}^{N}
   \left.\left(\frac{E_j^2\mbox{cosh}(\eta_t)}{2\;s\;E_t\;\mbox{cosh}^3(\eta_j)}\;\frac{d\sigma^{\scriptsize
           \mbox{NLO}}}{dR_2(J_{t},J_{j})}(m_t)
     \right)\right|_{\mbox{\scriptsize event } i}
\end{displaymath}
Fig. \ref{fig:MEM} shows the extraction of the top-quark mass with NLO (reddish) and Born likelihoods (bluish) with simultaneous variations of the renormalisation and factorisation scales by factors ${1\over2}$ (subscript) and $2$ (superscript) as a measure for higher order effects. The estimators are extracted as the minima of the negative logarithm of the likelihoods (`Log-Likelihoods') by a parabola fit.\\
\begin{minipage}{\linewidth}
\vspace{2ex}
 \centering 
    \includegraphics[width=0.49\textwidth]{{{%
          sgtsKT3-2ycut30membornnloincmu-100evts_pos-crop}}}
    \includegraphics[width=0.49\textwidth]{{{%
          sgttKT3-2ycut30membornnlomu-90evts_pos-crop}}}

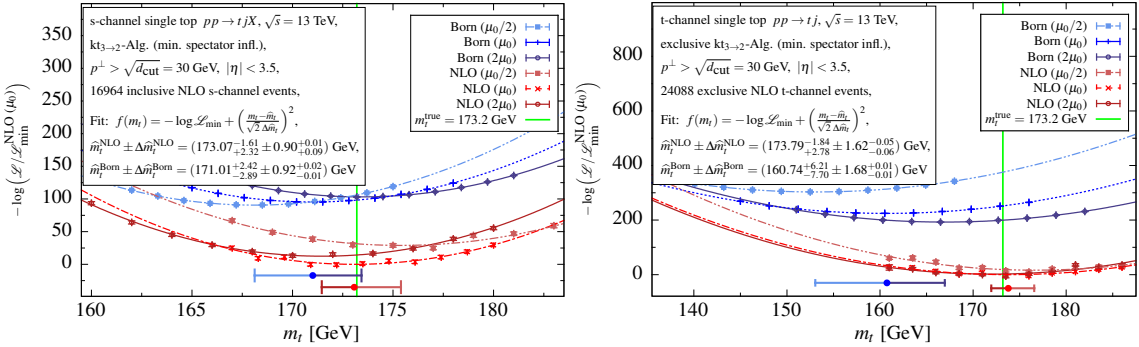
\captionof{figure}{Extraction of the top mass with MEM at NLO or Born accuracy.}
\label{fig:MEM}
\vspace{2ex}
\end{minipage}
The left hand side of fig. \ref{fig:MEM} shows that around $17000$
inclusive $s$-channel NLO events allow a precise mass determination with
a relative statistical uncertainty of
$\widehat{m}_t/\Delta\widehat{m}_t=0.5\%$. Applying the MEM at NLO
perfectly recovers the input value $m^{\scriptsize\mbox{true}}_t$.
However, using Born likelihoods on the same events introduces a bias
in the estimator $\widehat{m}_t^{\scriptsize\mbox{Born}}$ of $-2.2$
GeV. This bias is covered by the scale variations shifting the Born
estimator by $-2.9$ GeV and $+2.4$ GeV. The impact of the scale
variation on the NLO estimator is reduced to $-1.6$ GeV and $+2.3$
GeV. In contrast, the right hand side of fig. \ref{fig:MEM}
demonstrates that analysing NLO events with Born likelihoods can
result in large biases in the estimator which are not necessarily
covered by the scale variations. Applying the MEM at NLO to exclusive
$t$-channel NLO events again perfectly reproduces
$m^{\scriptsize\mbox{true}}_t$ while the Born estimator is shifted by
$-12.5$ GeV. Scale variations shift the Born estimator by $-7.7$ GeV
and $+6.2$ GeV. In the NLO analysis the impact of the scale variations
is significantly reduced to $-1.8$ GeV and $+2.8$ GeV.  In the Born
analysis the simultaneous variations of the renormalisation and
factorisation scales by factors ${1\over2}$ and $2$ do not give a
reliable estimate of the NLO effects. The large bias observed in the
Born estimator in the right hand side of fig. \ref{fig:MEM} would
require a significant calibration of the MEM which introduces related
uncertainties. It should be noted that the renormalisation scheme is
well-defined in the MEM at NLO. The extracted top-quark mass equates to the
pole mass. For a more in-depth study of single top-quark  production with the
MEM at NLO we refer to \cite{Martini:2017}.

\section{Conclusion}
We have presented an algorithm to extend the MEM to NLO
accuracy. We have illustrated the method using single top-quark
production as an example. We find a significant improvement compared
to leading-order predictions. In particular, the calibration required
when the MEM is applied in leading-order is reduced. Although not
illustrated here the method can also be used for new physics searches
using likelihood ratios.

\bibliographystyle{JHEP} 
\providecommand{\href}[2]{#2}\begingroup\raggedright\endgroup
\end{document}